\documentclass[12pt]{article}
\usepackage{color}
\usepackage{latexsym}
\usepackage{amsmath}
\usepackage{amsfonts}
\usepackage{color}
\usepackage{graphicx}
\usepackage{amssymb}
\usepackage{indentfirst} 
\numberwithin{equation}{section}
\newcommand{\be}{\begin{equation}}
\newcommand{\ee}{\end{equation}}

\title{Dynamics of spinning particles in pp-wave spacetimes} 
\author{
	\\ 
	\vspace*{0.4cm}
	K. Andrzejewski\footnote{
		University of  Lodz, Faculty of Physics and Applied Informatics. Poland, Lodz,  Pomorska 149/153, 	90-236; e-mail: krzysztof.andrzejewski@uni.lodz.pl}
	\vspace*{0.3cm}
}
\date{}
\begin{document}
\maketitle 
\begin{abstract} 
 In this work, we study the dynamics of a spinning particle in pp-wave spacetimes; in particular, plane gravitational waves and impulsive shockwaves. We present a step-by-step procedure that reduces the original problem to the dynamics of the transverse coordinates, in direct analogy with geodesic motion. To achieve this, we use an appropriate spin supplementary condition, Hamiltonian formulations (including a non-minimal one), and constants of motion associated with conformal Killing fields.
\end{abstract} 
\newpage 
\section{Introduction }
\label{s1}	
The motion of relativistic spinning bodies in external fields has been extensively studied for many years, providing valuable insights into spin–curvature coupling, precession effects, and corrections to gravitational-wave waveforms; see, e.g., Refs. \cite{b11a}-\cite{b10l} and the references therein. Such studies are important because realistic astrophysical scenarios typically involve rotating and extended objects, as well as strong-field effects. Recently, these investigations have gained additional motivation due to gravitational-wave observations, in particular those expected from the Laser Interferometer Space Antenna (LISA). However, analysing gravitational-wave emission from a spinning compact object inspiraling into a much more massive black hole (the so-called extreme mass-ratio inspiral, EMRI) requires the generation of accurate gravitational-wave templates that  account for spin effects \cite{bb1,bb2}. 
Therefore, analytical studies of the dynamics of spinning systems are of particular importance. However, the  spin–curvature coupling may destroy integrability and lead to chaotic behaviour; consequently, the systems preserving integrability  have attracted special attention \cite{b10j}-\cite{b10m}. These considerations provide the motivation and guiding principles  for the present study.
\par      
In the pole-dipole approximation, where the external field varies only weakly across the body, the dynamics can be reduced to a reference worldline $x^\alpha(\tau)$  together with   four-momentum $P^\alpha$ and spin tensor $S^{\alpha\beta}=-S^{\beta\alpha}$. These quantities are governed by the  Mathisson-Papapetrou-Dixon-Souriau   (MPDS)  equations  \cite{b1c}-\cite{b2e}\footnote{Our conventions: signature $(-,+,+,+)$, $x^\alpha{'}$ derivative w.r.t.  the $\tau$ parameter, $x'_\alpha=g_{\alpha\beta}x'^\beta$,  $D_\tau$ is the  covariant derivative of the tensor along the curve $x^\alpha(\tau)$.}  
\begin{align}
	\label{mpds1}
	{D_\tau P^\alpha}&=-\frac 12{R^\alpha}_{\beta\gamma\delta}S^{\gamma\delta}{x^\beta}{'}\, , \\
	\label{mpds2}
	{D_\tau S^{\alpha\beta}}&=P^\alpha{x^\beta}{'}-P^\beta{x^\alpha}{'}\, ,
	\end{align}
where $\tau$ is the proper time  parameter, i.e. $x^\alpha{'}x_\alpha{'}=-1$. 
\par 
The main difficulty in analysing the motion of a spinning body is that the MPDS equations alone are insufficient to uniquely determine both the reference trajectory and the spin evolution. To obtain a closed system, they have to be supplemented with an auxiliary constraint, the so-called spin supplementary condition (SSC), of the form
\be
\label{ssc}
S^{\alpha\beta} V_\alpha=0,
\ee
where $V$   is a vector field  defined, at least, along $x^\alpha(\tau)$.  Such a  constraint  can be interpreted as  the choice of   worldline requiring   the center of mass of the body measured by the observer with the velocity $V$ (thus $V$ should be time-like; or  at least, in the limiting case, null-like). 
\par 
The  choice of $V$ is a longstanding problem, and various conditions have been proposed; see \cite{b5a} for a wider discussion and references.
 Considerable effort has been devoted to understanding the role of different SSCs \cite{b3l}-\cite{b3mmmm}, in particular whether they merely correspond to different mathematical descriptions of the same physical system, in analogy with gauge transformations. On the other hand, alternative descriptions of spinning objects have also been developed, including Lagrangian, Hamiltonian, and other  effective formulations; see Refs. \cite{b11c}-\cite{b10g}.   
\par The  oldest and most  widely studied spin conditions are the Frenkel-Mathisson-Pirani (FMP)    spin condition $V^\alpha ={x^\alpha}{'}$  \cite{b1c,b1a,b3d},    and the       Tulczyjew-Dixon (TD) condition defined by   $V^\alpha=P^\alpha/M$  \cite{b2b,b2d,b3e}. Although each of these conditions has its own physical motivation (for example, through the definition of the mass:  $M^2\equiv-P_\alpha P^\alpha$  or  $ m\equiv -P_\alpha{ x^\alpha}{'}$, 
each of which is conserved only for particular choices of the SSC), due to the complex relation between momentum and velocity, they make the MPDS equations considerably more difficult to solve.  On the other hand, recently     Ohashi,  Kyrian and Semer\'ak  (OKS) \cite{b3h,b3i} have     proposed  a new condition.  Specifically, they proposed choosing  $V$  such that $D_\tau V=0$.  Then, it follows  that  the momentum $P^\alpha$ is proportional to the velocity, $P^\alpha =m x^\alpha{'}$, and $M=m$ is constant.  In consequence, the  MPDS equations \eqref{mpds1} and \eqref{mpds2} simplify to: 
\be  \label{e4} 
m D^2_\tau x^\alpha =-\frac 12{R^\alpha}_{\beta\gamma\delta}S^{\gamma\delta}x^\beta{'} ,
\quad \quad D_\tau  S^{\alpha\beta}=0.
\ee 
Thus, the OKS condition allows for a simpler analysis; moreover, in practice, it is equivalent to other spin supplementary conditions (see, e.g., \cite{b5a}). Consequently, it provides at least a qualitative description of the motion of spinning objects.
\par
Besides the MPDS equations, alternative approaches to the dynamics of spinning objects have been developed, namely the so-called effective theories.  They describe dynamics  of  point-like particles characterized by  an overall  position, momentum and spin. Usually, they   are  obtained in two ways: using the Lagrangian formalism  (see e.g. \cite{b11a,b11c,b3j} and references therein) or   by a  Hamiltonian description, see e.g. \cite{b10h,b10g,b10a,b10b,b10c,b10d,b10e,b4a}.       
 In latter case,  the starting point is an   effective Hamiltonian formalism in the extended phase space.  Namely,   first  we define the Poisson brackets:
 \be  \label{ep1}
\{x^\alpha,P_\beta\}=\delta^{\alpha}_{\beta}, \quad \{P_\alpha,P_\beta\}=-\frac{1}{2}R_{\alpha\beta\gamma\lambda}S^{\gamma\lambda} ,\quad \{S^{\alpha\beta},P_\gamma\}=\Gamma^\alpha_{\gamma\delta}S^{\beta\delta}- \Gamma^\beta_{\gamma\delta}S^{\alpha\delta},
\ee
 \be  \label{ep2}
\{S^{\alpha\beta},S^{\gamma\delta}\}=g^{\alpha\gamma}S^{\beta\delta}-g^{\alpha\delta} S^{\beta\gamma}+g^{\beta\delta}S^{\alpha\gamma}-g^{\beta\gamma}S^{\alpha\delta}.
\ee
Next,  a Hamiltonian should be defined. The most natural candidates are Hamiltonians for which the  resulting dynamics is equivalent to the MPDS equations supplemented by an appropriate spin supplementary condition.  
This approach was discussed in detail in Ref.  \cite{b4a} where appropriate Hamiltonians were constructed.     In particular, it  was   shown that  the dynamics generated by means of \eqref{ep1}, \eqref{ep2} and the  Hamiltonian 
 \be  \label{eh0}
H_0=\frac{1}{2m}g_{\alpha\beta} P^{\alpha}P^{\beta}=\frac{1}{2m}P^2,
\ee
  is    the same as for  MPDS equations together with  the OKS condition.
For the  FMP condition the dynamics is  governed by the Hamiltonian 
\be
\label{eh1}
H_1=\frac{1}{2m}\left( g^{\alpha\gamma} -\frac{1}{S^2}S^{\alpha\beta}{S^\gamma}_\beta \right) P_\alpha P_\gamma=H_0-\frac{1}{2mS^2}S^{\alpha\beta}{S^\gamma}_\beta P_\alpha P_\gamma,
\ee
where $S^2=S^{\alpha\beta}S_{\alpha\beta}/2$. 
\par 
For  the TD condition the Hamiltonian is  more involved   and can be found in Ref.  \cite{b4a}. 
The Hamiltonian approach described above  provides a new way to analyse the dynamics of spinning bodies.  Namely, motivated by various physical considerations (e.g. symmetry), one can  introduce alternative Hamiltonian formulations. An interesting example of such an approach  is a non-minimal Hamiltonian  proposed in Refs. \cite{b10a,b10b,b10c}, see also Sec. \ref{s6}.
\par
In  this work, we analyse the dynamics  of a   spinning body   in pp-wave spacetimes, in particular plane gravitational  waves and impulsive gravitational shockwaves. We pay special  attention on the  analytical considerations.   To this end, we employ the OKS condition, various Hamiltonian formalisms, and non-standard constants of motion; the latter  are our starting point. Namely, in Sec.  \ref{s2} we study  integrals of motion for  the MPDS equations  associated with conformal Killing fields.   Next, in Sec. \ref{s3}, we show that, for pp-waves and the OKS condition, the MPDS equations decouple, and their solution can be obtained through a step-by-step algorithm that reduces the problem to the dynamics of the transverse coordinates (similarly to the spinless case).  The obtained results are then applied to plane gravitational waves and illustrated with explicit examples  (Sec. \ref{s4})  as well as impulsive gravitational waves (Sec. \ref{s5}). Subsequently, in Sec. \ref{s6}, we analyse  the motion   when the Hamiltonian contains  a non-minimal  spin-spin  term.  
Finally,  in the last section we collect conclusions and   outline directions for future research.  
 \section{Conformal fields and spinning particles}
 \label{s2}
 It is well known that Killing vector fields give rise to integrals of motion and can therefore be used in the analysis of relativistic systems.  Remarkably, this observation has been extended to spinning particles  \cite{b12a}. A notable example is provided in Ref. \cite{bx0} where, for de Sitter spacetime and an arbitrary spin supplementary condition, the momentum and spin tensors are expressed explicitly in terms of  spacetime coordinates. On the other hand, it is also known that conformal Killing fields may also generate further constants of motion for spinless particles \cite{bx3}. Homothetic vector fields provide the simplest examples of this phenomenon; however, there exist spacetimes in which the same property also occurs for proper conformal Killing fields \cite{bx4}. In what follows, we investigate this issue in the context of spinning objects. The results obtained will subsequently be applied in Sec. \ref{s4}.
\par 
Let   us start with  the conformal Killing vector field $Y$, i.e. $\mathcal{L}_Yg=2\psi g$ where $\psi$  is  a function.  Then  using (skew)symmetry of the tensor $g_{\alpha\beta}$ ($S^{\alpha\beta}$, respectively) and the MPDS equations, after some calculations,  we obtain that   the quantity 
\be 
\label{e11a}
\tilde I=P^\alpha Y_\alpha-\frac 1 2 Y_{\alpha;\beta} S^{\alpha\beta},
\ee
 satisfies along the trajectory  the identity
\be
\label{e11}
\tilde I'= \psi  P^\alpha {x'_\alpha }+S^{\alpha\beta}x'_\beta\partial_\alpha\psi.
\ee
Obviously, for a Killing vector field (i.e. for $\psi=0$) one recovers the familiar integral of motion.
\par  
For the FMP condition we  have  $P^\alpha x_\alpha'=-m=const.$; moreover,  the last term  in eq. \eqref{e11} vanishes.  Consequently, we arrive at the   formula $\tilde I'=-m\psi$.
This result can also be derived within the Hamiltonian framework, see \eqref{ep1}, \eqref{ep2} and \eqref{eh1}. 
Namely,  after  straightforward but  tedious calculations, we get
\be
\label{e12}
\tilde I'=
\{\tilde I,H_{1}\}=\frac {1}{m} \psi P^\alpha (P_\alpha+\frac{1}{S^2}S_{\alpha\beta}{S^\beta}_\gamma P^\gamma)  +
\frac {1}{m}  S^{\delta\alpha}   (P_\alpha+\frac{1}{S^2}S_{\alpha\beta}{S^\beta}_\gamma P^\gamma)  \partial_\delta \psi.
\ee 
 Now, using the relation between momenta and velocities,  we obtain  the formula \eqref{e11}.  Thus, for  the FMP condition,  we get  the conserved quantity  $I=\tilde I+m\int \psi$. Obviously, this integral is  generally  non-local (it depends on the trajectory); 
 however,  similarly to the spinless case discussed in Ref.  \cite{bx3},   it can be shown that one can choose a parametrisation in which it becomes local.
 Moreover, in some special cases such  a localisation is possible for the  proper time parametrisation and then  $I$   reduces to an ordinary  constant of motion. For example,  this  holds for  a homothetic vector field $Y$, i.e. when $\psi=const.$; other examples, including  proper conformal Killing  fields, are provided by  pp-wave metrics, see \cite{bx4}.
 In view of the above, these results extend directly to spinning particles satisfying the FMP condition. Such conserved quantities may be useful because a generic pp-wave spacetime admits only one Killing vector field, whereas certain pp-wave geometries possess  three proper conformal Killing fields. Consequently, the associated charges are independent of those generated by Killing symmetries and therefore provide additional information about the dynamics. Finally, at the Hamiltonian level, we can compute the Poisson brackets of the localisable integrals of motion with the conserved quantities generated by Killing fields and obtain new localised charges. Consequently, the structure of the conformal algebra may play an important role in such considerations.
\par
For the  OKS condition  we have  $P^\alpha=mx'^\alpha$ and $m$  is  constant, thus  the first term in eq. \eqref{e11} is the same as in the previous case; however, wheres the second term,  in general, is not zero, making this case considerably more involved. This can be verified  by direct   calculations at  the Hamiltonian level 
 (see  eqs. \eqref{ep1}, \eqref{ep2} and \eqref{eh0}):  
\be 
\label{e13}
\tilde I'=
\{\tilde I,H_{0}\}=\frac {1}{m}(\psi P^2+S^{\alpha\beta}P_\beta\partial_\alpha\psi )=-\psi m +S^{\alpha\beta}x'_\beta \partial_\alpha\psi.
\ee
Integrating eq. \eqref{e13}, we get a generalized  integral of motion $I=\tilde I+\int (m\psi-S^{\alpha\beta}x'_\beta \partial_\alpha\psi)$.
Obviously, it localises  for homothetic fields. However, in Sec. \ref{s4},   we will show that    for some special classes of plane gravitational waves the integral term in  $I$  is  a function of $\tau$  and thus can be evaluated explicitly,  yielding a new independent integral of motion.
\section{The OKS condition and pp-waves}
\label{s3}
In this section,  we apply the OKS condition to a spinning body in  pp-waves.   We will work in the   Brinkmann  coordinates (they  will be useful in our further investigations of plane waves). 
\par  
Let us start with  a pp-wave metric\footnote{The indices  $a,b,\ldots$ run over  $1,2$; they are raising and lowering by $\delta_{ab}$.} 
\be
\label{e14}
g=2dudv+K(u,x^a)du^2+dx^adx_a.
\ee
The first equation in  \eqref{e4}  applied  for the $u$-coordinate gives  
$mu'=const.=p_v$;  thus  $u=p_v\tau/m$ can be used as an affine parameter. Next, the transverse part of the  geodesic equation is of the form  
\be
 \label{e15}
m{x^a}''=\frac{p_v^2}{2m}\partial_aK+\frac {p_v}{2m} \partial_a\partial_b K S^{bu}=\frac{p_v^2}{2m}\partial_a(K+\frac {1}{p_v} \partial_b K S^{bu} ) ,
\ee  
while  for  the $v$-coordinate we have 
\be
 \label{e16}
mv''=-\frac{p_v^2}{2m}\partial_u K-p_v\partial_aK x'^a+\frac{1}{2}\partial_a\partial_bK S^{ub}{x^a}{'}.
\ee
Equation \eqref{e16} can be integrated directly by using the velocity normalization condition:
\be
 \label{e17}
{x^a}' {x}_a'+2v'u'+(u')^2K=-1,
\ee
(this fact can be checked explicitly using the above   equations).
Next, for the OKS condition, the spin tensor  is parallel-transported, i.e.  ${D_\tau S^{\alpha\beta}}=0$,  thus we have the following equations  for the  spin tensor 
\begin{align}
	\label{e33a}
	{S^{ua}}'&=0,\\
	\label{e33b}
	{S^{ab}}{' }&=\frac{p_v}{2m}(\partial_a KS^{ub}+ S^{au}\partial_b K),\\
	\label{e33c}
	{S^{uv}}{'}&=\frac{p_v}{2m} S^{au}\partial_a K,
	\\
	\label{e33d}
	{S^{va}}{'}&= -\frac{p_v}{2m} S^{ba} \partial_b K+\frac{p_v}{2m} S^{vu}\partial_a K-\frac 1 2(K)'S^{ua}.
\end{align}
\par 
From the above discussion,  the solution procedure becomes clear. Namely, since $S^{ua}$ are  constants,    the  transverse dynamics, see eq. \eqref{e15},   decouples and takes a form analogous to the motion of a particle in an external potential. In general,  eq.  \eqref{e15}  may be difficult  to solve analytically;  similarly to the  geodesic motion of a spinless particle  \cite{bb4}, it may exhibit chaotic behaviour.  However,  for the specific  forms  of $K$,  it can be solved  exactly  as   will be shown in  the following sections. 
 Once the transverse motion is known, equations  for the  $v$-coordinate and  for $S^{ab}$,  $S^{uv}$ can be integrated. Having determined these quantities,  we  can integrate the last equation for $S^{av}$. Obviously,   the explicit integration  is   possible for  particular forms of  $K$, see e.g.  Sec. \ref{s4}. 
\par 
 The above procedure  can be  described in another (equivalent) way. To this end, let us  recall that   eqs. \eqref{e33a}-\eqref{e33d}   follow from  the OKS condition, i.e.  $S^{\alpha\beta}V_\beta=0$  with $D_\tau V=0$.     We    can therefore      write  $S^{uv}$ and $S^{bv}$  in terms   of the vector field $V$.
 Indeed,  assuming that  $V$ is time-like,   the component $V^u$ does not vanish.   Then, the condition \eqref{ssc}  for the  $uv$-component yields 
 \be
  \label{e18}
 S^{uv}=-\frac{1}{V^u}S^{ua}V^a,
 \ee
 while for the  $va$-component we get
 \be
  \label{19}
 S^{va}=\frac{1} {V^u}((KV^u+V^v)S^{au}+S^{ab}V^b).
 \ee
Thus, once the vector  field $V$ is known,   there is no need to integrate the equations governing   $S^{uv}$ and $S^{va}$.  
\par 
Moreover, the field $V$ can be reconstructed uniquely from its initial condition  $V(\tau_0)$. Namely, the $u$-component of the equation $D_\tau V=0$  yields  $V^u=const\neq 0$. Next,  $V^v$ can be  obtained from the  condition $V^2=-1$  provided we know $V^a$. For the latter components,  we have a  set of  differential  equations  
 \be
  \label{e20}
 {V^a}'=\frac{p_v}{2m} V^u\partial_aK.
 \ee  
 Thus,  eqs. \eqref{e33c} and \eqref{e33d} can be replaced  by  eq. \eqref{e20} (this can also  be  checked directly  by means of the above equations).
\section{Plane gravitational  waves and the  OKS condition}
\label{s4}
In this section, we consider the motion of a spinning body in plane-wave spacetimes. This problem  has been studied under various spin supplementary conditions. Namely,   the TD    condition has been investigated    in Refs. \cite{b13a,b13b}; however, due to  the complicated  relationship between momentum and velocity,  only partial results have been obtained. Recently, it has been shown that these difficulties can be avoided at linear order in spin  \cite{bxx}. On the other hand, the  FMP condition has been considered in Ref. \cite{bx1}; unfortunately,   a more detailed  analysis reveals that  the    solutions presented  do not satisfy the FMP  condition, so  this case  remains open.  Finally, an interesting  observation was made in Ref. \cite{b9}.  Namely, taking $V=\partial_v$, the authors obtain the  geodesic motion for the spinning body. 
Obviously, such a choice of $V$ may appear somewhat puzzling  (especially if one takes into account that $\partial_v$ is a null-like vector field).  However, in the previous work \cite{bx2} it was shown that such a field corresponds precisely to a special choice of the  OKS condition.  Here, following this idea   and using the results from Sec. \ref{s3},  we  will show  that   for an arbitrary  field satisfying the  OKS condition,  the dynamics of  spinning objects in plane gravitational waves admits a considerably simpler  description.
More precisely, the dynamics can be reduced to the spinless case,  which allows for the direct application of many results obtained for geodesic motion (see, for instance,  Ref. \cite{bb5}).
 Moreover,  we   show that the  integrals of motion associated with conformal Killing fields can  also be localised  in such spacetimes.       
\par
We shall work in Brinkmann coordinates, which are global and particularly convenient for  subsequent analysis.
So let us  take $K=K_{ab}(u)x^ax^b$  where $K_{ab}$\footnote{To simplify notation we omit the $u$-argument  in our considerations.  Moreover, note that  the vacuum solutions are characterized  by ${K^a}_a=0$ (sometimes, they are called exact plane waves).  } is a symmetric matrix. Due to the relation between $u$ and $\tau$, we  simplify the  notation and work in the  $u$-parametrisation -  dot  will refer to the  derivative w.r.t. to $u$. In our case eq. \eqref{e15} reduces to     
\be
 \label{e21}
\overset{..}{x}^a=K_{ab}(x^b -S^{ub}/p_v).
\ee
Thus,  defining 
\be 
\label{e21a}
{ y}^a=x^a -S^{ua}/p_v,
\ee 
eq. \eqref{e21} takes exactly the same form as in the spinless case, i.e. the transverse coordinates are shifted only by   constants. 
Consequently,  the integrability of the transverse equation is preserved. 
\par 
Remarkably, the equation governing the $v$ coordinate can be integrated explicitly.  Namely,  by straightforward calculations, we verify that the following function:  
\be
 \label{e22}
2v=- \dot x^a x_a +\frac{S^{uv}}{p_v}-\frac{m^2}{p_v^2}u+const.
\ee
satisfies  eq. \eqref{e17}. This result is a direct consequence of the symmetry structure of plane gravitational waves discussed in Sec. \ref{s2}. Namely,  the  field $Y=2v\partial_v+x^a\partial_a$ is a homothetic vector field with the conformal factor $\psi=1$. Thus, we can apply results from  Sec. \ref{s2} (see eq. \eqref{e13}). As a consequence, we obtain an additional integral of motion
\be
 \label{e23}
I=2mvu'+mx^a x'^a-S^{uv}+m\tau, 
\ee
taking into account that  $\tau=um/p_v$ we   immediately get  eq. \eqref{e22}.
\par  
Now, we consider the spin components. Let us start with $S^{uv}$. Using  eq. \eqref{e21} we   rewrite eq. \eqref{e33c}   in the form
\be
 \label{e24}
\dot S^{uv}=-\overset{..}{x}^aS^{ua}-K_{ab}S^{ua} S^{ub}/p_v.
\ee 
In consequence, we  get the solution: 
\be
\ \label{e25}
S^{vu}=\dot x^aS^{ua} +\hat K_{ab}S^{ua} S^{ub}/p_v,
\ee
where $ {\hat  K_{ab}}=\int { K_{ab}}=\int K_{ab}(u)du$. 
In view of eqs.  \eqref{e22} and   \eqref{e25}, the $v$-coordinate has the same integrability properties as in the spinless case.
Similarly,    using  eq.  \eqref{e21}  we solve   eq. \eqref{e33b}, the final result reads 
\be
 \label{e26}
S^{ab}=\dot x^aS^{ub}-\dot x^b S^{ua}+ \frac{1}{p_v}(S^{ub}\hat K_{ad}S^{ud}-S^{ua}\hat K_{bd}S^{ud}).
\ee
It remains  to  integrate eq. \eqref{e33d}. This is a slightly more involved task.  Using eqs. \eqref{e33b} and \eqref{e33c}, after straightforward but lengthy calculations, we get that  
\be
 \label{e27}
S^{av}=\frac{1}{2}KS^{ua}+\frac{1}{2}S^{au}(\dot x^b+S^{uc}\hat K_{bc} /p_v)(\dot x^b+S^{ud}\hat K_{bd} /p_v)-S^{ab}(\dot x^b+S^{uc}\hat K_{bc} /p_v),  
\ee 
is a  solution to eq. \eqref{e33d}. 
\par 
In view of the above, under the OKS condition, the entire dynamics of a spinning particle is encoded in the transverse equation \eqref{e21} (which, in turn, can be reduced to its spinless counterpart; see eq. \eqref{e21a}) and in the computation of the integrals defining the matrix $\hat K_{ab}$.  Thus, solutions of the geodesic equations play a central role in describing the spin evolution. In particular, exact solutions allow one to determine the spin dynamics explicitly under the OKS condition;  examples illustrating this point are provided in the next section.
\par
At the end, it is  worth  rewriting the   dynamics of the  spin tensor by  introducing vector $z^a$ of the  following form
\be
 \label{e28}
z^b=(\dot x^b+S^{uc}\hat K_{bc} /p_v).
\ee 
Then,   we obtain a more transparent representation of the spin components
\be
 \label{e29}
S^{vu}=z^aS^{ua},\quad S^{ab}=S^{ub}z^a-S^{ua}z^b, \quad S^{av}=\frac 12 KS^{ua}+\frac 12 S^{ua}z^bz^b-S^{ub}z^az^b,
\ee   
which   simplifies   the  analysis of  spin dynamics. 
\par  
Finally, note that taking $V=\partial_v$ in the OKS condition, we have   $S^{ua}=0$ and  we recover the results of Ref. \cite{b9}, i.e. the geodesic motion and $S^{\alpha\beta}$ are constant.   
\subsection{Some examples}
\label{s4a}
The simplest case  of  plane waves is given by the constant matrix $K_{ab}$;  to  be specific, let us consider a vacuum solution $K_{ab}=\omega^2 \textrm{diag}(1,-1)_{ab}$.  Then, solutions  to eq. \eqref{e21} are given by    $x^1=A_1\cosh(\omega u)+B_1\sinh(\omega u)+S^{u1}/p_v$ and $x^2=A_2\cos(\omega u)+B_2\sin(\omega u)+S^{u2}/p_v$ and  $\hat K_{ab}=K_{ab}u$. From the above discussion, we  immediately obtain a complete description of the  dynamics (for all coordinates  and spin tensor). 
\par Next,  let us consider the so-called  sandwich wave, see \cite{bx5}, or  other profiles described in Ref.  \cite{bb5}.  Then,  the transverse  equation  \eqref{e21} and $\hat K_{ab}$ can  also be    found; in consequence, the  entire dynamics as well.  However, we do not discuss these cases further.  Instead,  we study an   example of a   continuous pulse.   It   will be useful    for our further  studies of  impulsive gravitational shockwaves, and will illustrate  the  issue  of the integrals of motion associated with conformal Killing fields presented in  Sec. \ref{s2}.  
\par
Namely, let us take the pulse of plane gravitational wave defined by the profile 
\be
 \label{e30}
K_{ab}=\frac{\kappa}{(u^2+\epsilon^2)^2}\textrm{diag(1,-1)}_{ab}\ ,
\ee 
Then,   the solutions of transverse equations, for  $\kappa<\epsilon^2$,   read
\be
 \label{e31}
x^a(u)=C_1^a \sqrt{u^2+\epsilon^2}\sin({\Lambda_a}\tan^{-1}( {u}/{\epsilon})+C_2^a)+S^{ua}/p_v,
\ee
where $ \Lambda _a=\sqrt{1+(-1)^a\frac{\kappa}{\epsilon^2}}$. 
\par Before we go further let us make a comment. Imposing the conditions $\dot x^a(-\infty)=0$  and next $x^a(-\infty)=x^a_{in}$ we obtain 
\be
 \label{e32}
 x^a(u)=\frac{(x^a_{in}-S^{ua}/p_v)}{\epsilon \Lambda_a} \sqrt{u^2+\epsilon^2}\sin({\Lambda_a}(\tan^{-1}( {u}/{\epsilon)+\pi/2}))+S^{ua}/p_v,
\ee
In consequence,  the final velocity  is   $\dot x^a(\infty)=\frac{\sin({\Lambda_a}\pi) }{\epsilon{\Lambda_a}}  (x^a_{in}-S^{ua}/p_v)$, thus it depends on the spin tensor. However, the cross-section, which is   related to the  derivative of  $x^a_{in}$ with respect to $\dot x^b(\infty)$, see  \cite{bx5}, does not depend on  spin, and thus is the same as for the spinless case (this coincides  with the  transformation \eqref{e21a}).
\par   
Next,  we  find that  in our case the matrix $\hat K_{ab}$ is also  given  explicitly:
\be
\label{e33}
\hat K_{ab}=\frac{\kappa}{2\epsilon^3} \left(\frac{\epsilon u}{u^2+\epsilon^2}+\tan^{-1}(u/\epsilon)\right)\textrm{diag}(1,-1)_{ab} ,
\ee 
Having the exact forms of $x^a$  and $\hat K_{ab}$ we can easily find $v$ and all remaining spin components, see equations in  \eqref{e29}.
\par   
The above results allow us to analyse  the change in the spin components  after the passage of the wave (a type of  spin memory effect). To this end, let us note that $\hat K_{ab}(\pm\infty)=\pm\frac{\kappa\pi}{4\epsilon^3}\textrm{diag}(1,-1)_{ab}$; thus, assuming that  the  particle  is at rest at $u=-\infty$,  we have  $z^a(-\infty)=-\frac{\kappa\pi}{4\epsilon^3p_v}(S^{u1},-S^{u2})$, while 
$z^a(\infty)=\dot x^a(\infty)+\frac{\kappa\pi}{4\epsilon^3p_v}(S^{u1},-S^{u2})$.  In consequence, the change of spin components, $S^{\alpha\beta}(\infty)-S^{\alpha\beta}(-\infty)$, can also be directly  computed, see eqs. \eqref{e29}.
\par 
At the end of this  section,  we will illustrate the procedure for the   integrals of motion described in Sec. \ref{s2}.  To this end let us note  that the plane wave metric with the profile \eqref{e30} exhibits a  conformal Killing field: 
\be
\label{e34a} 
Y=Y^\alpha\partial_\alpha =(u^2+\epsilon^2)\partial_u-\frac 1 2 x^ax_a\partial_v+ux^a\partial_a,
\ee
 with the conformal factor $\psi=u$.  Remarkably,  the conformal factor depends on $u$ only, on the other hand    $u$ is proportional to $\tau$, thus we can  use   results from  Sec. \ref{s2} 
to find an additional local   integral of motion. 
\par  
First, we find that  
\be
\label{e34} 
\tilde I=P^\alpha Y_\alpha -uS^{vu}-\left(\frac{\kappa}{u^2+\epsilon^2}-1\right)x^1S^{u1}+\left(\frac{\kappa}{u^2+\epsilon^2}+1\right)x^2S^{u2}.
\ee
Thus,  under the FMP condition,  eq. \eqref{e11}   immediately  implies that   the quantity   $I=\tilde I+p_v\tau^2/2=\tilde I+m^2u^2/p_v$  is a constant of motion (obviously we still  need the velocity-momentum relation, see e.g.  \cite{b4a}). 
For the OKS condition studied in this section, the  situation is more involved; however, due to the   direct relation between momenta and velocities, the interpretation of the resulting constant of motion is more transparent.  Namely,  for  the OKS condition,  we have an additional term  in eq.  \eqref{e13}.  For    $Y$  given by    \eqref{e34a} we obtain    
 \be
 \label{e35}
  \dot {\tilde I}= -\frac{m^2}{p_v}u-(S^{vu}+S^{au}\dot x^a).
 \ee 
 Using eq. \eqref{e25}  we get
 \be 
 \label{e36}
 \dot {\tilde I}= -\frac{m^2}{p_v}u - \frac {1}{p_v}\hat K_{ab}S^{ua}S^{ub}.
\ee
Now, we can easily integrate  both side  of eq. \eqref{e36}; as a result  we obtain  the following integral of motion
\be 
 \label{e37}
I=\tilde I+\frac{m^2}{2p^v}u^2+\frac{1}{p_v}S^{ua}S^{ub}\int\hat  K_{ab},
\ee
where 
\be
 \label{e38}
\int \hat K_{ab}=\frac{\kappa}{2\epsilon^3}u\tan^{-1}(u/\epsilon)\textrm{diag}(1,-1)_{ab}.  
\ee
Using     $\tilde I$  given by eq. \eqref{e34}, after some computations,   we arrive at the final  form of the integral of motion 
\be
 \label{e39}
I=P^\alpha Y_\alpha+\frac{m^2}{2p_v}u^2 -u\dot x^aS^{ua}-((S^{u1})^2-(S^{u2})^2)\frac{\kappa u^2}{2p_v\epsilon^2(u^2+\epsilon^2)}  +x^aS^{ua}+\frac{\kappa}{u^2+\epsilon^2}(S^{u2}x^2-S^{u1}x^1),
\ee 
where $P^\alpha=mx'^\alpha=p_v\dot x^\alpha $ and $Y_\alpha$  are determined  by \eqref{e34a}. 
\par 
To analyse    $I$, let us recall that the transverse part of the equations of motion  is directly related to the spinless case (see eq. \eqref{e21a}). On the  other hand, we have seen that the remaining dynamical variables expressed in terms of the transverse variables and  $\hat K_{ab}$.       In view of this,  we  can expect that  a similar situation holds for $I$. To see this, first,   we  express the velocity $\dot v$ appearing  in $I$  by means of eq. \eqref{e23}. Then $I$ becomes, apart from $u$ and $S^{ua}$,  a function of the transverse  components $x^a$ and their derivatives. After performing the transformation \eqref{e21a} in  $I$, after some computations,    we observe that  $I$ reduces to the spinless expression   presented in  \cite{bx6}; i.e. the sum of two Lewis invariants.  They, in turn,  directly yield the solutions \eqref{e31};  this closes  the picture   of the  transverse dynamics  of the spinning particle in the profile \eqref{e30}. 
\par 
A similar situation also holds  for another example of  plane gravitational wave. It is defined by a more  complicated profile $K_{ab}$ and  consequently its polarization is not linear; making it physically more interesting.  Namely, let  us consider a plane gravitational wave defined by  
\be
K_{ab}=\frac{\kappa}{(u^2+\epsilon^2)^2}
\left(
\begin{array}{cc}
	\cos(\phi(u)) &\sin(\phi(u))\\
	\sin(\phi(u)) & -\cos(\phi(u))
\end{array}
\right),
\ee
where 
\be
\phi(u)=\frac{2\gamma}{\epsilon}\tan^{-1}(u/ \epsilon),
\ee 
It turns out that, see \cite{bb3}, such a gravitational  wave has the non-homothetic   conformal vector field of the form    
\be
\label{n1}
\overline {Y}=Y+\gamma(x^1\partial_2-x^2\partial_1),
\ee  
where $Y$ is given by \eqref{e34a} and the conformal factor  is the same as above ($\psi=u$).
In this case, we can also easily  find   the matrix $\hat K_{ab}$:
\be 
\hat K_{11}=-\hat K_{22}=\frac{\kappa(2\gamma u\cos(\phi(u))+(\epsilon^2-2\gamma^2+u^2)\sin(\phi(u)))}{4\gamma(\epsilon^2-\gamma^2)(u^2+\epsilon^2)} ,
\ee
\be 
\hat K_{12}=\hat K_{21}=\frac{\kappa(2\gamma u\sin(\phi(u))-(\epsilon^2-2\gamma^2+u^2)\cos(\phi(u)))}{4\gamma(\epsilon^2-\gamma^2)(u^2+\epsilon^2)} .
\ee
 On the other hand, using the results of \cite{bx6} and eq. \eqref{e21a}, we  can explicitly  find   (even in terms of elementary functions) the general solution of the equation \eqref{e21}. Thus,  as in the previous example, the evolution of the $v$-coordinate, as well as the spin tensor, can be obtained explicitly, see eqs. \eqref{e28} and \eqref{e29}.  Moreover,  due to   following relation 
 \be
 \hat K_{ab}(\pm \infty)=\frac{\kappa}{4\gamma(\epsilon^2-\gamma^2)}
 \left(
 \begin{array}{cc}
 	\pm\sin(\pi\gamma/\epsilon) &-\cos(\pi\gamma/\epsilon)\\
 	-\cos(\pi\gamma/\epsilon) &\mp\sin(\pi\gamma/\epsilon)
 \end{array}
 \right),
 \ee
 we can also find the  memory effect related to the spin. 
 \par 
   Now, let us return to conformal symmetry and integrals of motion. Since the metric describes a plane gravitational wave and conformal factor depend on $u$ only, we  infer that the   integral of motion associated with the field $\bar Y$  localises. In addition, it can be explicitly determined. Indeed, substituting   the conformal filed 
\eqref{n1}  into eq. \eqref{e11a} we easily get $\tilde I$  for our case;  next, by direct calculations we  find  that  the matrix elements $\int \hat K_{ab}$ take the form 
\be 
\int \hat K_{11}=-\int \hat K_{22}=\frac{\kappa(\gamma \cos(\phi(u))+u\sin(\phi(u)))}{4\gamma(\epsilon^2-\gamma^2)}, 
\ee
\be 
\int\hat K_{12}=\int\hat K_{21}=\frac{\kappa(\gamma \sin(\phi(u))-u\cos(\phi(u)))}{4\gamma(\epsilon^2-\gamma^2)} .
\ee    
Substituting  both $\tilde I$ and $\int \hat K_{ab}$ into eq. \eqref{e37} we obtain the final form of  the integral of motion associated with the conformal field $\overline {Y}$.
\par  Finally, let us recall that the conformal factor depends only on the $u$-coordinate for all N-type pp-waves, see  \cite{bx4} and references therein. Thus for such spacetimes and the FMP condition the integrals of motion associated with conformal fields localise and can provide new information (as we discussed in Sec. \ref{s2}).  For the OKS condition, additional information about the solutions of the transversal equation is required.  
\section{Impulsive gravitational shockwaves}
\label{s5}
The impulsive gravitational shockwaves are pp-waves with  $\delta(u) $ profile. More precisely, they are described by the metric \eqref{e14} with a function $K$  of the form
\be
 \label{e40}
K=f(x^1,x^2)\delta(u).
\ee   
The famous example is the Aichelburg-Sexl metric with  $f(x^1,x^2)\propto \ln(r) $; it describes the gravitational field generated by an ultrarelativistic  particle. The second important example is the impulsive plane gravitational wave defined by  
$K(x^1,x^2)\propto\delta(u)((x^1)^2-(x^2)^2)$. 
\par For such spacetimes, even in the spinless case, a mathematically rigorous analysis of the dynamics is rather subtle \cite{bx7}.
 Thus, for  a spinning particle,  we restrict ourselves to some general  observations.
To this end, let us observe that  by setting  $\kappa=4\epsilon^3/\pi$ in the profile \eqref{e30} and then taking the limit   $\epsilon\rightarrow 0 $, we obtain an impulsive gravitational wave. Now,  taking  the same limit in the solutions  obtained in Sec. \ref{s4a},  we   find the dynamics of a spinning particle in the impulsive  plane  gravitational wave.  Namely, we find   that $x^a$    exhibits a  velocity jump at $u=0$. On the other hand,  the matrix $\hat K_{ab}$ tends to the step function $\theta(u)$.  Thus, the   dynamics of  $z^a$ (see eq. \eqref{e28})     is  governed by the step function of the  $u$ coordinate. Consequently, the spin tensor also exhibits a discontinuity across the wavefront, i.e. $u=0$  (see eq. \eqref{e29}).
\par 
For an arbitrary impulsive shockwave  (described by a function $f$)  the transverse equations take the form    
\be
 \label{e41}
\overset{..}{x}^a=\frac{\delta(u)}{2}\partial_a(f-\frac{1}{p_v}S^{ub}\partial_b f), 
\ee 
thus  $x^a=x_0^a+v_1^au$ for $u<0$ and $x^a=x_0^a+v_2^au$ for $u>0$,  i.e. they remain continuous across the impulse, whereas their derivatives undergo a finite jump. This jump can be  determined  by integrating the corresponding equation  \eqref{e41}:  
\be 
v_2^a-v_1^a=\frac{1}{2}\partial_a(f-\frac{1}{p_v}\partial_bfS^{ub})(x_0).
\ee
The velocity discontinuity therefore depends on the spin components $S^{ua}$. As in the plane impulsive wave considered above, we also find a discontinuity in the spin dynamics.

\section{Non-minimal Hamiltonian}
\label{s6} 
The Hamiltonians introduced in Sec. \ref{s1} may provide a starting point for alternative descriptions of spinning particles.   Such an approach was adopted, e.g., in Ref. \cite{b10a,b10b,b10c} where a non-minimal extension  of the Hamiltonian $H_0$ was proposed.   Specifically, motivated by the gravitational analogue of the Stern-Gerlach interaction,   a curvature-mediated spin-spin interaction was introduced:         
\be
 \label{e42}
H_2=H_0+\frac{\kappa}{4}R_{\alpha\beta\gamma\delta}S^{\alpha\beta}S^{\gamma\delta}=\frac {P^2}{2m}+\frac{\kappa}{4}R_{\alpha\beta\gamma\delta}S^{\alpha\beta}S^{\gamma\delta}.
\ee
A remarkable feature of this extension is that the integrals of motion associated with $H_2$ coincide with those of $H_0$ \cite{b10b,b10c}. 
We also note that the relation $P^\alpha=mx'^\alpha$ is still  valid. However, substituting  this relation  into the  Hamiltonian \eqref{e42}  we find  that  $x'_\alpha x'^\alpha$  is not  constant; thus $\tau$ is  not, in general, the proper-time parameter.
\par 
The non-minimal part of $H_2$ implies  additional  terms in the equations of motion. Namely,  we have to add $-\frac \kappa 4 S^{\epsilon\beta}S^{\gamma\delta}\nabla^\alpha R_{\epsilon \beta\gamma\delta}  $  in  eq.   \eqref{mpds1}. 
While for  the spin tensor we get
\be
 \label{e43}
D_\tau S^{\alpha\beta}=
\kappa S^{\gamma\delta}({R^\alpha}_{\epsilon\gamma\delta}S^{\beta\epsilon}-{R^\beta}_{\epsilon\gamma\delta}S^{\alpha\epsilon}).
\ee
To gain further insight into the physical implications of the proposed modification,  we investigate the dynamics in various gravitational backgrounds.   Such investigations have been carried out for black hole spacetimes in Ref.  \cite{b10e}. Here, we extend these investigations to pp-wave spacetimes.
\par 
Before doing so, we first consider de Sitter spacetime, or more generally, spacetimes of constant curvature. Then,  the  non-minimal  term is proportional to $S^2$ and thus the dynamics of $x^\alpha$ and $S^{\alpha\beta}$ is the same as for $H_0$. 
\par
Now,  let us consider pp-wave spacetimes. Then for the metric  \eqref{e14} we obtain
\be 
 \label{e44}
H_2=\frac{1}{2m}P^2-\frac \kappa 2 \partial_a\partial_bKS^{ua}S^{ub}.
\ee 
Using the Hamiltonian $H_2$ together with the Poisson brackets \eqref{ep1} and \eqref{ep2} we find that $u'$ and $S^{ua}$ are constant (as in the OKS case discussed in Sec. \ref{s3}).    
Equation for $v$ is extended by the term $\frac \kappa 2\partial_u\partial_a\partial_b K S^{ua}S^{ub}$. Nevertheless, it can still be integrated straightforwardly by means of the Hamiltonian constraint:
 \be
 \label{e44a}
 m (2v'u'+Ku'^2+x'^a x'_a)-\kappa \partial_a\partial_bK S^{ua}S^{ub} =2H_2=const.
 \ee
Finally, the transverse equations take the form 
\be
 \label{e45}
m{x^a}{''}=\frac{p_v^2}{2m}\partial_aK-\frac {p_v}{2m} \partial_a\partial_b K S^{ub}  +\frac \kappa 2 S^{ub}S^{uc} \partial_a\partial_b\partial_cK.
\ee   
Thus, in general, the transverse dynamics is modified by the non-minimal term.  However, for plane gravitational waves, $K=K_{ab}x^ax^b$, the non-minimal term disappears and the transverse dynamics coincides with that discussed in  Sec. \ref{s4}.  
In summary, for pp-wave spacetimes,  in particular for plane waves, the form of the coordinate dynamics is very similar to that obtained under the OKS condition. 
\par
Now, let us focus on the spin dynamics. For simplicity, we restrict our attention to plane waves  (this case is particularly interesting because  for the OKS condition, i.e. $H_0$, the equation governing the $v$-coordinate can be integrated exactly). We therefore consider  a plane gravitational wave, i.e. $K=K_{ab}x^ax^b$. Then, the equation for  $S^{uv}$ reduces  to 
\be
S'^{uv}=u'K_{ab}x^bS^{au}+2\kappa K_{ab}S^{ua}S^{ub}.
\ee      
Since  for the plane waves the transverse dynamics is the same as in Sec. \ref{s4}, we can use  eq. \eqref{e21} to integrate the above equation; the final result (in the $u$ parametrisation) reads  
\be 
\label{e46}
S^{uv}=-\dot x^a S^{ua}-\frac{1-2\kappa m}{p_v}\hat K_{ab}S^{ua}S^{ub},
\ee
comparing eqs. \eqref{e25}   and \eqref{e46},  we observe that the non-minimal coupling only modifies the coefficient appearing in the last term (translates  the numerator factor).
\par  
Now, let us consider $S^{ab}$. Then, the non-minimal correction takes the form  $2\kappa S^{ud}(K_{bd}S^{ua}-K_{ad}S^{ub}) $. Again, using the transverse equation we obtain  
\be
\label{e47}
\dot S^{ab}=\overset{..}{x}^aS^{ub}-\overset{..}{x}^bS^{ua} +\frac{1-2m\kappa}{p_v}S^{ud}(K_{ad}S^{ub}-K_{bd}S^{ua}). 
\ee
The above equation can be integrated straightforwardly, and the result is as in eq. \eqref{e26}  with the replacement  $1/p_v\rightarrow\frac {1-2\kappa m}{p_v}$, i.e. exactly as in the case of   $S^{uv}$.  Finally, let us consider $S^{av}$. First, we find  the corresponding  equation of motion 
\be
\label{e48}
S^{av}{'}= \frac{p_v}{m} K_{bc}x^cS^{ba}-\frac{p_v}{m} K_{ac}x^cS^{vu}+(K)'S^{ua}+2\kappa( K_{ab}S^{ub}S^{vu}+K_{cb}S^{ac}S^{ub}).
\ee
Now,  using eqs. \eqref {e45}, \eqref{e46} and \eqref{e47}   after  tedious computations,  we obtain   that $S^{av}$ is given by eq. \eqref{e27}  with the same replacement $1/p_v\rightarrow\frac {1-2\kappa m}{p_v}$. 
\par 
Finally, let us  return to the $v$-coordinate. Using  Hamiltonian $H_2$, one finds, after some calculations,   that eq. \eqref{e44a} can be integrated; namely,  we arrive at   the following form of the $v$-coordinate 
\be
\label{e49}
 2vu'+x'^a x^a-\frac{S^{uv}}{m}-\frac{2\tau}{m} H_2=const. 
\ee   
 In summary,  for  spinning particles   in   plane gravitational waves     the dynamics  of  $u,S^{ua}$  is the same under the  OKS condition (i.e. governed by  $H_0$) and  for  the  non-minimal Hamiltonian $H_2$; in particular, these quantities do not depend on $\kappa$.  The components $S^{uv},S^{ab},S^{av}$ retain the same functional form upon the replacement     $1/p_v\rightarrow\frac {1-2\kappa m}{p_v}$. A similar statement applies for the $v$-coordinate; however, let us stress that   $S^{uv}$ in eq.  \eqref{e49} depends on $\kappa$. 
 \par 
Finally,   taking $\kappa=1/2m$ we see that  the  equations simplify  considerably.  Interestingly, this observation is fully consistent\footnote{
Let us note that a different convention is used in Ref. \cite{b10b}, leading to $\kappa=-1/(2m)$ there.
} with the result obtained in Ref. \cite{b10b}. Namely, for such a choice of $\kappa$   the FMP condition is preserved  up to terms of cubic order in spin.  
 \section{Conclusions and outlook} 
 \label{s8} Let us summarise our results. In this work, we investigated the dynamics of spinning objects in pp-wave spacetimes. Using the OKS condition (or, equivalently, an appropriate Hamiltonian framework), we showed that the equations of motion can be treated through a step-by-step procedure. Moreover, we demonstrated that the vector field defining the OKS condition can be used to reconstruct the spin components. 
 A crucial ingredient of this procedure  is the solvability of the transverse equations of motion; this, in turn, is closely related to the solvability of the corresponding geodesic equations. Consequently, the method is particularly effective for plane gravitational waves, where the transverse dynamics is governed by a linear system. For certain  wave profiles, this leads to exact analytical solutions and enables a direct description of the spin dynamics, in particular after the passage of the wave. In the general case, however, chaotic behaviour may arise, similarly to the spinless case.
\par 
 Furthermore, in the case of plane gravitational waves, we observed a significant role played by  integrals of motion associated with conformal Killing fields. In particular, homothetic vector fields facilitate the integration of the equation for the $v$-coordinate, whereas proper conformal Killing fields may assist in solving the transverse equations. We also applied our approach to impulsive gravitational shockwaves. In particular, for impulsive plane waves we employed a limiting procedure based on smooth plane-wave profiles. In such spacetimes, we found discontinuities in both the velocities and the spin evolution.
 \par 
 We also investigated the dynamics generated by a non-minimal Hamiltonian containing a spin-curvature-spin interaction term. In particular, we showed that for plane gravitational waves the equations can still be treated systematically and that, compared with the OKS case, the effect of the non-minimal coupling essentially amounts to a modification of the parameters entering the spin components. For a particular choice of the coupling constant, the equations simplify considerably.
\par
 Regarding possible extensions, it would be worthwhile to investigate the role of conformal Killing fields in other classes of pp-wave spacetimes, see for example \cite{bx4,bx6}, as well as for non-minimal Hamiltonian models. The analysis of non-minimal Hamiltonian models itself deserves further study, especially concerning the distinguished coupling values discussed here and in Ref.~\cite{b10b}. Another interesting direction would be to extend the present analysis to spacetimes with non-vanishing torsion \cite{b15b}, or to problems related to gravitational deflection and lensing \cite{b15c}.
 \par 
 Finally, let us recall that charged spinning particles in electromagnetic backgrounds can also be described within the pole-dipole approximation. This problem has been investigated using various approaches; see \cite{b5a,b4mmm} for reviews. It turns out that  the equations governing the electromagnetic case differ qualitatively from their gravitational counterparts. For example, the evolution parameter does not necessarily coincide with the proper time. 
 A the same time, numerous analogies between gravitational and gauge theories have been explored in recent years. A notable example is the classical double-copy correspondence, which may be viewed as a classical manifestation of colour-kinematics duality; see, among others, \cite{b7a,b7b,b7c}. Within this framework, certain gravitational phenomena can be understood in terms of gauge theories. In particular, pp-wave spacetimes correspond to specific abelian gauge configurations. More precisely, the electromagnetic counterpart of a pp-wave geometry is characterised by the potential
 $
 A_{\tilde u}=A(\tilde u,\tilde x^a)= \frac{p_v}{2q}K,
$
where $K$ denotes the profile function appearing in the pp-wave metric \eqref{e14} and  ($\tilde u,\tilde v,\tilde x^a$) denote light-cone coordinates in Minkowski spacetime.
 \par 
Motivated by this correspondence, as well as by previous studies of spinless particles \cite{b8a,b8b,b8c} and  gravito-electromagnetic analogies \cite{b6b}, it would be interesting to compare the dynamics of spinning particles in gravitational and electromagnetic backgrounds related through the double-copy prescription. Such an analysis  requires a more comprehensive discussion of the theory of charged spinning bodies and therefore lies beyond the main scope of the present work. Nevertheless, preliminary investigations suggest that the dynamics of spinning particles satisfying the OKS condition in pp-wave spacetimes closely resembles that in electromagnetic fields, provided that an appropriate spin supplementary condition is imposed in the latter case, see  \cite{bx2}.
 \par
 In view of these electromagnetic considerations, it would also be interesting to investigate the OKS condition within the framework of   Kaluza-Klein geometry, which provides a unified description of gravitational and electromagnetic interactions; in analogy with the analysis for the  FMP condition presented in Ref. \cite{b15a}.
\vspace{0.5cm}
\par
{\bf Acknowledgments}
\par
The author is grateful to the anonymous referee for  valuable comments  and suggestions.


\begin{thebibliography}{99}  
	
\bibitem{b11a}R. Porto, ``Post-Newtonian corrections to the motion of spinning bodies in nonrelativistic general relativity" Phys. Rev. D {\bf 73}  (2006)  104031
\bibitem{b3k}S. Hergt, J. Steinhoff,  G. Sch\"afer,  ``The reduced Hamiltonian for next-to-leading-order spin-squared dynamics of general compact binaries" Class. Quant.   Grav. {\bf 27} (2010)  135007

\bibitem{b3m}R. Porto, ``Next to leading order spin-orbit effects in the motion of inspiralling compact binaries"  	Class. Quant. Grav. {\bf 27} (2010) 205001 
\bibitem{b10h}
F. Blanco, E. Flanagan, ``Motion of a spinning particle under the conservative piece of the self-force is Hamiltonian to first order in mass and spin"
Phys. Rev. D {\bf 107} (2023) 124017 
\bibitem{b10i}
K. Wang, C.-J. Feng, ``Spin vector deviation and the gravitational wave memory effect between two free-falling gyroscopes in the plane wave spacetime" Phys. Rev. D {\bf 107} (2023) 084044
\bibitem{b10k}
S. Jumaniyozov, J. Rayimbaev, Y.  Turaev,
``Dynamics of spinning particles around a charged black-bounce spacetime"
    Eur. Phys. J. C {\bf 85} (2025)  1247
\bibitem{b10l}
Y. Chen, K. Wang, C.-J. Feng,
``Higher order analysis of the gravitational wave velocity memory effect between two free-falling gyroscopes in the plane wave spacetime" Phys. Rev. D {\bf 111} (2025) 104085
\bibitem{bb1}
N. Warburton, T. Osburn, C.  Evans, ``Evolution of small-mass-ratio binaries with a spinning secondary" Phys. Rev. D {\bf 96} (2017) 084057 
\bibitem{bb2}
V. Skoup\'y, G. Lukes-Gerakopoulos, ``Spinning test body orbiting around a Kerr black hole: Eccentric equatorial orbits and their asymptotic gravitationa	l-wave fluxes" Phys. Rev. D {\bf 103} (2021)  104045
\bibitem{b10j}
V. Witzany,  G. Piovano,  
``Analytic Solutions for the Motion of Spinning Particles near Spherically Symmetric Black Holes and Exotic Compact Objects"   Phys. Rev. Lett. {\bf 132} (2024)  171401
\bibitem{b3o}P. Ramond,  ``On the integrability of extended test body dynamics around black holes"     Class. Quant. Grav.  {\bf 42} (2025) 065019 
\bibitem{b10m}V. Skoup\'y, V. Witzany,  ``Analytic Solution for the Motion of Spinning Particles in Kerr Spacetime"  Phys. Rev. Lett. {\bf 134}  (2025) 171401

\bibitem{b1c}M. Mathisson, ``Neue mechanik materieller systemes"  Acta Phys. Polon. {\bf 6} (1937) 163 
\bibitem{b2a}A. Papapetrou, ``Spinning test particles in general relativity"  Proc. R. Soc. London, Ser. A  {\bf 209} (1951) 248
\bibitem{b2b} W.  Dixon, ``A Covariant Multipole Formalism for Extended Test Bodies in General Relativity" Nuovo Cimento {\bf 34} (1964) 317
\bibitem{b2d} W.  Dixon, ``Dynamics of extended bodies in general relativity. I. Momentum and angular momentum" Proc. R. Soc. London, Ser. A {\bf 314} (1970)  499
\bibitem{b2e} J. Souriau, ``Mod\'ele de particule \'a spin dans le champ \'electromagn\'etique et gravitationnel" Ann. I. H. P. {\bf 20} (1974)  315
\bibitem{b5a}
L. Costa, J. Nat\'ario,  ``Center of Mass, Spin Supplementary Conditions, and the Momentum of Spinning Particles" in Equations of Motion in Relativistic Gravity  Springer (2015) 215 

\bibitem{b3l} J. Steinhoff, ``Canonical Formulation of Spin in General Relativity"  Ann. Phys. {\bf 523} (2011) 296
\bibitem{b3mm}G. Lukes-Gerakopoulos, J. Seyrich,  D. Kunst, ``Investigating spinning test particles: spin supplementary conditions and the Hamiltonian formalism" Phys.  Rev. D {\bf 90}  (2014) 104019
\bibitem{b3mmm}
G.  Lukes-Gerakopoulos,  ``Time parameterizations and spin supplementary conditions of the Mathisson-Papapetrou-Dixon equations"  	Phys. Rev. D {\bf 96} (2017)  104023
\bibitem{b3n}
L.  Costa, G.  Lukes-Gerakopoulos, O. Semer\'ak, ``Spinning particles in general relativity: Momentum-velocity relation for the Mathisson-Pirani spin condition" Phys. Rev. D {\bf 97} (2018)  084023 
\bibitem{b3mmmm}
I. Timogiannis, G. Lukes-Gerakopoulos, T.  Apostolatos, ``Spinning test body orbiting around a Schwarzschild black hole: Comparing Spin Supplementary Conditions for Circular Equatorial Orbits" Phys. Rev. D {\bf 104}  (2021)  024042

\bibitem{b11c}A. Deriglazov, W. Guzm\'an Ram\' irez, ``Lagrangian formulation for Mathisson-Papapetrou-Tulczyjew-Dixon (MPTD) equations"  Phys. Rev. D {\bf 92} (2015) 124017
\bibitem{b6b}L. Costa, J.  Nat\'ario, M.  Zilh\~ao, ``Spacetime dynamics of spinning particles - exact electromagnetic analogies"  Phys. Rev. D {\bf 93}  (2016)  104006 
\bibitem{b4mmm} A.  Deriglazov, W.  Guzm\'an Ram\' irez    ``Recent progress on the description of relativistic spin: vector model of spinning particle and rotating body with gravimagnetic moment in General Relativity"  Adv.  Math.  Phys.   {\bf 2017} (2017) ID 7397159 	 
\bibitem{b10g}V.  Witzany,  ``Hamilton-Jacobi equation for spinning particles near black holes" 	Phys. Rev. D {\bf 100} (2019)  104030 

\bibitem{b1a}	J. Frenkel,  ``Die Elektrodynamik des rotierenden Elektrons" Z. Phys. {\bf37} (1926) 243
\bibitem{b3d}F. Pirani, ``On the Physical significance of the Riemann tensor"  Acta Phys. Pol. {\bf 15} (1956) 389 

\bibitem{b3e}W. Tulczyjew, ``Motion of multipole particles in General Relativity theory"  Acta  Phys. Pol. {\bf 18} (1959) 393 



\bibitem{b3h}A. Ohashi, ``Multipole particle in relativity"  Phys. Rev. D {\bf  68} (2003) 044009 
\bibitem{b3i}K. Kyrian, O. Semer\'ak, ``Spinning test particles in a Kerr field II"  Mon. Not. Roy. Astron. Soc. {\bf 382} (2007) 1922



\bibitem{b3j} E. Barausse, E. Racine,  A. Buonanno, ``Hamiltonian of a spinning test-particle in curved spacetime" Phys. Rev. D
{\bf 80}  (2009) 104025


\bibitem{b10a}
I. Khriplovich, ``Particle with internal angular momentum in a gravitational field" Sov. Phys. JETP {\bf 69} (1989) 217 
\bibitem{b10b}
G. d’Ambrosi, S. Satish Kumar, J. van Holten, “Covariant hamiltonian spin dynamics in curved spacetime"  Phys. Lett. B {\bf 743} (2015) 478 
\bibitem{b10c} J. van Holten, ``Spinning bodies in General Relativity" Int. J. of Geom. Methods in Mod. Physics {\bf 13} (2016) 1640002


\bibitem{b10d}
R. Rietdijk, J. van Holten, ``Spinning particles in Schwarzschild space-time" Class. Quant. Grav. {\bf 10} (1993) 575 

\bibitem{b10e}G. d’Ambrosi, S. Satish Kumar, J. van de Vis,  J.  van Holten, ``Spinning bodies in curved spacetime" Phys. Rev. D {\bf 93}  (2016) 044051


\bibitem{b4a}V. Witzany, J.  Steinhoff,  G.  Lukes-Gerakopoulos, ``Hamiltonians and canonical coordinates for spinning particles in curved space-time" 	 Class. Quant.  Grav. {\bf 36} (2019) 075003 


\bibitem{b12a}R. R\"udiger, ``Conserved Quantities of Spinning Test Particles in General Relativity. I"
Proc. Roy. Soc. Lond.  A {\bf 375} (1981) 185 



\bibitem{bx0}
Y. Obukhov, D. Puetzfeld, ``Dynamics of test bodies with spin in de Sitter spacetime" Phys. Rev. D {\bf 83} (2011) 044024 
\bibitem{bx3}
N. Dimakis, P. Terzis, T. Christodoulakis, ``Integrability of geodesic motions in curved manifolds through non-local conserved charges" Phys. Rev. D {\bf 99} (2019) 104061
\bibitem{bx4}
 M. Elbistan, N. Dimakis,  K. Andrzejewski,  P. Horvathy, P. Kosi\'nski, P.-M. Zhang  ``Conformal symmetries and integrals of the motion in pp waves with external electromagnetic fields" 
Ann. Phys. {\bf  418} (2020) 168180 
\bibitem{bb4} 
J. Podolsk\'y,  K.  Vesel\'y, ``Chaos in pp-wave spacetimes" Phys. Rev. D {\bf  58} (1998) 081501(R)
\bibitem{b13a}M. Mohseni, R.  Tucker, C. Wang, ``On the motion of spinning test particles in plane gravitational waves" 	Class. Quant. Grav. {\bf 18} (2001) 3007
\bibitem{b13b}S. Kessari, D. Singh, R. Tucker, C. Wang, ``Scattering of Spinning Test Particles by Plane Gravitational and Electromagnetic Waves" 	Class. Quant. Grav. {\bf 19} (2002) 4943
\bibitem{bxx}K. Wang ``Analytic Solution for the Motion of Spinning Particles in Plane Gravitational Wave Spacetime" arXiv:2601.21438 (2026)

\bibitem{bx1}
M. Mohseni, “Spinning particles in gravitational wave spacetime” Phys. Lett. A {\bf 301} (2002) 382


\bibitem{b9}D. Bini, G. Gemelli, ``Scattering of spinning test particles by gravitational plane waves" Il  Nuovo Cimento B {\bf 112} (1997) 165

\bibitem{bx2}K. Andrzejewski,  ``Revisiting the dynamics of a charged spinning body in curved spacetime" Class. Quant.  Grav. {\bf 42} (2025) 055019
\bibitem{bb5}
J. Podolsk\'y, K. Vesel\'y, ``New examples of sandwich gravitational waves and their impulsive limit" Czech. J. Phys. {\bf 48} (1998) 871
\bibitem{bx5}
J.  Garriga, E.  Verdaguer,  ``Scattering of quantum particles by gravitational plane waves" Phys. Rev. D {\bf 43} (1991) 391
\bibitem{bx6}
K. Andrzejewski, S. Prencel, ``Niederer's transformation, time-dependent oscillators and polarized gravitational waves"  Class. Quant.  Grav. {\bf 36} (2019) 155008
\bibitem{bb3}B. Tupper, A. Keane, G. Hall, A. Coley, J. Carot,``Conformal symmetry inheritance in null fluid spacetimes" Class. Quant. Grav. {\bf 20 }(2003) 801
\bibitem{bx7}R. Steinbauer,  
``Geodesics and geodesic deviation for impulsive gravitational waves"
J. Math. Phys. {\bf 39} (1998) 2201

\bibitem{b15b}
S. Hojman, ``Lagrangian theory of the motion of spinning particles in torsion gravitational theories"  Phys. Rev. D {\bf  18} (1978) 2741
\bibitem{b15c}
Z. Zhang, G. Fan, J. Jia,  ``Effect of Particle Spin on Trajectory Deflection and Gravitational Lensing"  JCAP {\bf 09} (2022) 061


\bibitem{b7a}R. Monteiro, D. O'Connell, C. White, ``Black holes and the double copy" JHEP {\bf 12} (2014) 056
\bibitem{b7b}N. Bahjat-Abbas, A. Luna, C.  White,  ``The Kerr-Schild double copy in curved spacetime" JHEP {\bf 12} (2017) 004
\bibitem{b7c} M.  Gurses, B.  Tekin,  ``Classical Double Copy: Kerr-Schild-Kundt metrics from Yang-Mills Theory"  Phys. Rev.  D {\bf 98}  (2018)  126017 

\bibitem{b8a}
I. Bia\l ynicki-Birula,  S. Charzy\'nski, ``Trapping and
Guiding Bodies by Gravitational Waves Endowed with
Angular Momentum" Phys. Rev. Lett. {\bf 121} (2018) 171101 
\bibitem{b8b}A. Ilderton, ``Screw-symmetric gravitational waves: a double copy of the vortex" Phys. Lett. B {\bf 782} (2018) 22  
\bibitem{b8c}K. Andrzejewski, S. Prencel, ``From polarized gravitational waves to analytically solvable  electromagnetic beams" Phys. Rev. D {\bf 100}  (2019) 045006 


\bibitem{b15a}
F. Cianfrani, I. Milillo, G. Montani,
``Dixon-Souriau equations from a 5-dimensional spinning particle in a Kaluza-Klein framework"  	Phys. Lett. A {\bf 366} (2007) 7













\end{thebibliography}
\end{document}